\documentstyle[aps,epsfig]{revtex}                  
\begin{document}
\pagestyle{empty}                                      
\preprint{
\font\fortssbx=cmssbx10 scaled \magstep2
\hbox to \hsize{
\hfill$\raise .5cm\vtop{
                \hbox{NCTU-HEP-0103}}$}}
\draft
\vfill
\title{On Analytic Properties of the Photon Polarization Function in a
Background Magnetic Field  }

\author{W. F. Kao, Guey-Lin
Lin and Jie-Jun Tseng} %

\address{Institute of Physics, National Chiao Tung University,
Hsinchu 300, Taiwan}
\date{\today}
%
%
\vfill
\maketitle
\begin{abstract}
We examine the analytic properties of the photon polarization
function in a background magnetic field, using the technique of
inverse Mellin transform. The photon polarization function is
first expressed as a power series of the photon energy $\omega$
with $\omega< 2m_e$. Based upon this energy expansion and the
branch cut of the photon polarization function in the complex
$\omega$ plane, we compute the absorptive part of the polarization
function with the inverse Mellin transform. Our results are valid
for arbitrary photon energies and magnetic-field strengths. The
applications of our approach are briefly discussed.

\end{abstract}
%
%
%
%
\pagestyle{plain}

The behavior of a charged particle in a background magnetic field
is rather well known. The energy of the charged particle is
quantized according to the Landau levels. For a non-relativistic
electron moving in a uniform magnetic field along the $+z$
direction, its energy levels are given by
\begin{equation}
E_{n,s_z}=\left(n+{1\over 2}+s_z \right)\omega_c+{p_z^2\over
2m_e},
\end{equation}
where $\omega_c\equiv  eB/m_e$ ($e>0$) is the cyclotron frequency
of the electron and $s_z$ is the electron spin projection along
the $+z$ direction. For a relativistic electron (positron), the
energy quantization becomes
\begin{equation}
E^2_{n,s_z}=m_e^2+p_z^2+eB\left(2n+1+2s_z\right),
\end{equation}
with the following correspondent wave function
\begin{equation}
\psi^{\pm}_{n,p_y,p_z,s_z}(\vec{r},t)=
\exp(-iE_n t+ip_y y+ip_z z){\bf F}^{\pm}_{n,p_z,s_z}(x'),
\end{equation}
where ${\bf F}^{\pm}_{n,p_z,s_z}(x')$ is a four-component spinor
with $x'=x-p_y/eB$. The detailed form for ${\bf F}^{\pm}_{n,p_z,s_z}$
can be found, for example, in\cite{JL}.

While the wave function of a charged fermion in the background
magnetic field is well understood, it is often rather non-trivial
to compute a physical process occurring in a background magnetic
field, particularly, if there are more than one charged fermions
involved in the reaction. A famous example is the pair production
process $\gamma\to e^+ e^-$, which is relevant to the gamma-ray
attenuation in the neutron star\cite{STU}, and was first studied
by Toll\cite{toll} and Klepikov\cite{kle} independently. The
approach by Toll and Klepikov is based upon a direct squaring of
$\gamma\to e^+ e^-$ matrix elements using exact electron and
positron wave functions in a background magnetic field. The most
updated calculation using this approach can be found in \cite{DH}.
To understand this problem better, however, one should demonstrate
that the pair production width can also be obtained from the
absorptive part of photon polarization function in a background
magnetic field. In this regard, Tsai and Erber\cite{TE} obtained
the absorptive part of the one-loop photon polarization function
in the asymptotic limit $\omega\gg 2m_e$ and $B\ll B_c\equiv
m_e^2/e$. Their result was shown\cite{TE} to agree with that of
Toll and Klepikov. However, in the aforementioned asymptotic
limit, the threshold behavior of the pair-production width is
completely absent. Later on Shabad\cite{shabad} obtained the
absorptive part (and the dispersive part as well) of one-loop
photon polarization function for a general photon energy and
magnetic-field strength. In that work, the threshold behavior of
the pair-production width was worked out explicitly. We remark
that Refs.\cite{TE,shabad} employed Schwinger's proper-time
representation for the electron and positron Green's
functions\cite{sch} inside the photon polarization functions. To
our knowledge, Ref.\cite{shabad} is the first work which shows
that the proper-time representation for the photon polarization
function gives equivalent pair-production width to that obtained
from squaring the $\gamma\to e^+ e^-$ amplitude directly.
Unfortunately, the manipulations of Ref.\cite{shabad} are rather
involved and the substantial details of them were given in some
other unpublished preprints\cite{prep}. It is not very clear how
one can generalize the approach of Ref.\cite{shabad} to other
processes.

In this work, we will provide an alternative derivation of the
pair-production width (or equivalently the absorptive part of the
photon polarization function) from the proper-time representation
of the photon polarization function. We shall also outline the
procedure of obtaining the dispersive part of the polarization
function, which is relevant to the photon index of refraction. It
will be clear that our approach is very straightforward and
physically intuitive. Furthermore, it is applicable to many
processes occurring in a constant background electromagnetic
field. In fact, there are growing activities on computing the
two-point and three-point current correlation functions in a
constant background field, using the string-inspired world-line
formalism\footnote{For an overview on recent developments, see
\cite{ds,schubert}.}. The final results of these calculations are
expressed in terms of multiple integrals similar to the
proper-time representations. Our approach will be very useful for
investigating the analytic properties of these integrals, hence
the analytic behaviors of various current correlation functions.

To illustrate our approach, let us begin with the proper-time
representation of photon polarization function $\Pi_{\mu\nu}$ in a constant
background magnetic field\cite{TS}:
\begin{eqnarray}
\Pi_{\mu\nu}(q)&=&-{e^3B\over (4\pi)^2}\int_0^{\infty}ds
\int_{-1}^{+1} dv \{e^{-is\phi_0}[(q^2g_{\mu\nu}-
q_{\mu}q_{\nu})N_0 \nonumber \\
&-&(q_{\parallel}^2g_{\parallel\mu\nu}-
q_{\parallel\mu}q_{\parallel\nu})N_{\parallel}
+(q_{\bot}^2g_{\bot\mu\nu}-
q_{\bot\mu}q_{\bot\nu})N_{\bot}]\nonumber \\
&-&e^{-ism_e^2}(1-v^2)(q^2g_{\mu\nu}-
q_{\mu}q_{\nu})\},
\label{proper_t}
\end{eqnarray}
where $q$ is the photon four-momentum with $q_{\parallel}^{\mu}\equiv (\omega, 0, 0, q_z)$ and
$q_{\bot}^{\mu}\equiv (0, q_x, q_y, 0)$ for
a magnetic field in the $+z$ direction,
\begin{equation}
\phi_0=m_e^2-{1-v^2\over 4}q_{\parallel}^2-{\cos(zv)-\cos(z)\over 2z\sin(z)}
q_{\bot}^2
\end{equation}
with $z=eBs$, and
\begin{eqnarray}
N_0&=&{\cos(zv)-v\cot(z)\sin(zv)\over \sin(z)},\nonumber \\
N_{\parallel}&=&-\cot(z)\left(1-v^2+{v\sin(zv)\over \sin(z)}\right)
+{\cos(zv) \over \sin(z)},\nonumber \\
N_{\bot}&=&-{\cos(zv)\over \sin(z)}
+{v\cot(z)\sin(zv)\over \sin(z)}+2{\cos(zv)-\cos(z)\over \sin^3(z)}.
\end{eqnarray}
The two independent eigenmodes of the above polarization tensor
are $\epsilon^{\mu}_{\parallel}$ and $\epsilon^{\mu}_{\bot}$ which
are respectively parallel and perpendicular to the plane spanned
by the photon momentum ${\bf q}$ and the magnetic field ${\bf B}$.
They obey the eigenvalue equations
$\epsilon^{\mu}_{\parallel}(-q^2g_{\mu\nu}+\Pi_{\mu\nu})
\epsilon^{\nu}_{\parallel}\equiv q^2+\Pi_{\parallel}=0$ and
$\epsilon^{\mu}_{\bot}(-q^2g_{\mu\nu}+\Pi_{\mu\nu})
\epsilon^{\nu}_{\bot}\equiv q^2+\Pi_{\bot}=0$ respectively with
$\Pi_{\parallel,\bot}=\epsilon^{\mu}_{\parallel,\bot}
\Pi_{\mu\nu}\epsilon^{\nu}_{\parallel,\bot}$. It turns out that
$\Pi_{\parallel}$ and $\Pi_{\bot}$ are proportional to
$N_{\parallel}$ and $N_{\bot}$ respectively. We shall not discuss
the contribution by $N_0$ since it does not correspond to an
independent eigenmode. The absorptive part of
$\Pi_{\parallel,\bot}$ gives rise to the photon absorption
coefficient (pair-production width) via the relation
$\kappa_{\parallel,\bot}={\rm Im} \Pi_{\parallel,\bot}/\omega$.

To study the analytic properties of Eq.~(\ref{proper_t}), we
employ the sum rule developed by us\cite{klt}:
\begin{equation}
{1\over n!}\left({d^n\over d(\omega^2)^n}
\Pi_{\parallel,\bot}\right)\Big{\vert}_{\omega^2=0}
={M_{\parallel,\bot}^{1-2n}\over
\pi}\int_{0}^{1}dy_{\parallel,\bot}\cdot y_{\parallel,\bot}^{n-1}
\cdot \left(\kappa_{\parallel ,\bot}
y_{\parallel,\bot}^{-1/2}\right). \label{mellin}
\end{equation}
where $M_{\parallel}$ and $M_{\bot}$ are threshold energies for
pair productions with $M_{\parallel}^2-q_z^2=4m_e^2$ and
$M_{\bot}^2-q_z^2=m_e^2(1+\sqrt {1+2B/B_c})^2$;
$y_{\parallel,\bot}=M_{\parallel,\bot}^2/\omega^2$. One notes that
the absorptive part of $\Pi_{\parallel,\bot}(\omega^2)$ vanishes
for the range $0 \le \omega^2 \le M^2_{\parallel,\bot}$\cite{adl}.
Therefore one can effectively set the integration range of Eq.
(\ref{mellin}) as from $y_{\parallel,\bot}=0$ to
$y_{\parallel,\bot}=\infty$. Now, it is easily seen that the
derivatives of $\Pi_{\parallel,\bot}$ at the zero energy are
proportional to the Mellin transform of $\kappa_{\parallel,\bot}
\cdot y_{\parallel,\bot}^{-1/2}\equiv \kappa_{\parallel,\bot}\cdot
\omega/M_{\parallel,\bot}$. Once the l.h.s. of Eq. (\ref{mellin})
is calculated, the absorption coefficients
$\kappa_{\parallel,\bot}$ or ${\rm Im}\Pi_{\parallel,\bot}$ can be
determined by the inverse Mellin transform.

In order to compute the derivatives of $\Pi_{\parallel,\bot}$ at
the zero energy, we note that a rotation of the integration contour
$s\to -is$ is permissible for a photon energy below the pair-production
threshold.
This rotation turns the oscillating trigonometric
functions in $\Pi_{\parallel,\bot}$ into more manageable hyperbolic
functions. In the limit that $B\ll B_c$, one can perform an asymptotic
expansion in $B/B_c$, which gives
\begin{equation}
{1\over n!}\left({d^n\over d(\omega^2)^n}
\Pi_{\parallel,\bot}\right)\Big{\vert}_{\omega^2=0}
={2\alpha_e m_e^2\over \pi}\left({B^2\sin^2\theta\over 3B_c^2m_e^2}\right)^n
{\Gamma(3n-1)\Gamma^2(2n)\over \Gamma(n)
\Gamma(4n)}\left({6n+1, 3n+1\over 4n+1}\right)+\cdots,
\label{diff}
\end{equation}
where $\theta$ is the angle between directions of the photon
momentum and the magnetic field, and the neglected terms are
higher order in $B/B_c$. Applying the inverse Mellin transform, we
obtain absorption coefficients $\kappa_{\parallel,\bot}$ in
agreement with previous results by Tsai and Erber\cite{TE}. Such
results are however not satisfactory. As one can easily see, apart
from the trivial mass factor $m_e^{2-2n}$ , the first term on the
R.H.S. of Eq.~(\ref{diff}) grows to infinity as $n$ increases, no
matter how small the ratio $B/B_c$ is. This implies that the
disregarded higher-order terms are in fact non-negligible for a
sufficiently large $n$. As a result, the large moments of
$\kappa_{\parallel,\bot}$ are not accurately determined by the
first term of the above asymptotic expansion. Hence the threshold
behaviors of $\kappa_{\parallel,\bot}$ are not seen after applying
the inverse Mellin transform. To generate correct threshold
behaviors for absorption coefficients, we shall not expand in the
magnetic-field strength even if $B\ll B_c$. We will obtain photon
absorption coefficients valid for arbitrary photon energies and
magnetic-field strengths.

It is simpler to analyze $\Pi_{\parallel,\bot}$ in the special case that
$q_z=0$, i.e., ${\bf q} \cdot {\bf B}=0$.
Without losing the generality, we may further choose $q^\mu
=(\omega,q_x,0,0)$. It is easy to show that
$\Pi_{\parallel}=-(\alpha_e\omega^2/ 4\pi) \bar{\Pi}_{\parallel}$
and  $\Pi_{\bot}=-(\alpha_e q_x^2/ 4\pi) \bar{\Pi}_{\bot}$ with
\begin{eqnarray}
\bar{\Pi}_{\parallel}(\omega^2,q_x^2)&=&
\int _0 ^\infty dz \int _{-1}^{+1} dv \; \exp [ {-is \phi_0}]
N_{\parallel}, \nonumber \\
\bar{\Pi}_{\bot}(\omega^2,q_x^2)&=&
\int _0 ^\infty dz \int _{-1}^{+1} dv \; \exp [ {-is \phi_0}]
N_{\bot}.
\label{pibar}
\end{eqnarray}
Clearly, for the general case that $q^{\mu}=(\omega,q_x,0,q_z)$
with $q_z^2=q_x^2\cot^2\theta$, the photon polarization functions,
denoted as $\hat{\Pi}_{\parallel,\bot}(\omega^2,q_x^2,\theta)$,
are given by $\hat{\Pi}_{\parallel}(\omega^2,q_x^2,\theta)=
-(\alpha_e\omega^2\sin^2\theta/ 4\pi)
\bar{\Pi}_{\parallel}(q_{\parallel}^2,q_x^2)$, with
$q_{\parallel}^2\equiv \omega^2-q_z^2=\omega^2-q_x^2\cot^2\theta$;
and $\hat{\Pi}_{\bot}(\omega^2,q_x^2,\theta)=-(\alpha_e q_x^2/
4\pi) \bar{\Pi}_{\bot}(q_{\parallel}^2,q_x^2)$. Let us now return
to the special case. Rotating the integration contour $s\to -is$
and performing the energy expansion in $\omega$, we arrive at
\begin{equation}\label{component}
\Pi_{\parallel}(\omega^2,q_x^2)=-\frac{\alpha_e \omega^2}{4\pi}
\left( A(\alpha_a,\beta_a) +B(\alpha_b,
\beta_b)+B(\alpha_b^{\prime},\beta_b^{\prime})+C(\alpha_c,\beta_c)
+C(\alpha_c^{\prime},\beta_c^{\prime})\right),
\end{equation}
and
\begin{equation}\label{pi_perp}
\Pi_{\perp}(\omega^2,q_x^2) = \frac{\alpha_e q_x^2}{4\pi} \left(
A(\alpha_a, \beta_a)+
C(\alpha_c,\beta_c)+C(\alpha_c^{\prime},\beta_c^{\prime})+
D(\alpha_d,\beta_d)\right),
\end{equation}
where the functions $A$, $B$, $C$ and $D$ are defined
as\footnote{For convenience, we shall suppress the subscripts of
$\alpha$ and $\beta$ except in some special cases.}
\begin{eqnarray}\label{Nzv}
A(\alpha, \beta) &=&
  {\sum}K^A_{lmpp'}(q^{\prime 2})
    (\frac{1}{B^{\prime}})^n \frac{({\omega^{\prime}}^2)^{n-l}}{(n-l)!}
\int _0 ^\infty dz \; z^{n-l} \;
    \exp [ -z\beta] \int _{0} ^{1} dv \; (1-v^2)^{n-l}
\left(\exp [ \alpha  zv]+\exp [ -\alpha  zv]\right), \\
B(\alpha, \beta) &=& - {\sum} K^B_{lmpp'}(q^{\prime 2})
    (\frac{1}{B^{\prime}})^n \frac{({\omega^{\prime}}^2)^{n-l}}{(n-l)!}
   \int _0 ^\infty dz \; z^{n-l} \;
    \exp [ -z\beta] \int _{0} ^{1} dv \; (1-v^2)^{n-l+1}
\left(\exp [ \alpha zv]+ \exp [ -\alpha zv]\right),\\
C(\alpha, \beta) &=& -
 {\sum} K^C_{lmpp'}(q^{\prime 2})
    (\frac{1}{B^{\prime}})^n \frac{({\omega^{\prime}}^2)^{n-l}}{(n-l)!}
   \int _0 ^\infty dz \; z^{n-l} \;
    \exp [ -z\beta]
\int _{0} ^{1} dv \; v(1-v^2)^{n-l}
\left(\exp [ \alpha zv]-\exp [ -\alpha zv]\right),\\
D(\alpha, \beta) &=&-
  {\sum} K^D_{lmpp'}(q^{\prime 2})
    (\frac{1}{B^{\prime}})^n \frac{({\omega^{\prime}}^2)^{n-l}}{(n-l)!}
\int _0 ^\infty dz \; z^{n-l} \;
    \exp [ -z\beta] \int _{0} ^{1} dv \; (1-v^2)^{n-l}
\left(\exp [\alpha  zv]+\exp [-\alpha  zv]\right),
\end{eqnarray}
with ${\sum} \equiv \sum_{n=0}^{\infty} \sum_{l=0}^n
\sum_{m=0}^{\infty}\sum_{p',p=0}^l$, $\omega^{\prime 2}\equiv
\omega^2/4m_e^2$, $B'\equiv B/B_c$, $q^{\prime 2}\equiv
q_x^2/4m_e^2$,
\begin{eqnarray}\label{kfunction}
K^A_{lmpp'}(q^{\prime 2})&=& \frac{2(-1)^{l+p+p^{\prime}}\,
(2q^{\prime 2})^l \Gamma(l+m+1)}{
(l-p)!p!(l-p^{\prime})!p^{\prime}! \Gamma(m+1)}, \ \ \
K^B_{lmpp'}(q^{\prime 2})=\frac{1}{2}K^A_{lmpp'}(q^{\prime 2}), \nonumber \\
K^C_{lmpp'}(q^{\prime 2})&=& K^A_{lmpp'}(q^{\prime
2})\left(\frac{l+m+1}{l+1}\right), \ \ \ K^D_{lmpp'}(q^{\prime
2})=\frac{8 (-1)^{l+p+p^{\prime}}\, (2q^{\prime 2})^l
\Gamma(l+m+3)(l+1)}{ (l+1-p)!p!(l+1-p^{\prime})!p^{\prime}!(l+2)
\Gamma(m+1)}.
\end{eqnarray}
The actual arguments in the functions $A$, $B$, $C$ , and $D$ are
given by $\alpha_a=\alpha_c=\alpha_c^{\prime}=p'-p+1$,
$\alpha_b=\alpha_b^{\prime}=\alpha_d=p'-p$, $\beta_a=
p+p'+2m+1+1/B'$, $\beta_b=\beta_a-1$,
$\beta_b^{\prime}=\beta_a+1$, $\beta_c=\beta_a$,
$\beta_c^{\prime}=\beta_a+2$, and $\beta_d=\beta_a+1$. Note that
the various indices in the summation arise as follows: the index
$n$ comes from the photon-energy expansion, $l$ arises from the
binomial expansion of $((1-v^2)\omega^{\prime
2}+2(\cosh(zv)-\cosh(z))q^{\prime 2}/z\sinh(z))^n$, $p$ and $p'$
arises from writing $(\cosh(zv)-\cosh(z))^l$ as a sum of
exponential functions, and $m$ is due to expansions such as $\sinh
^{-l} z = 2^l \exp [-lz] \sum^\infty_{m=0} C_m^{l+m-1}\exp
[-2mz]$. It may appear at a first glance that our expansion on
$\sinh ^{-l} z$ and other similar terms are not convergent at
$z=0$. For $z\in (0,\epsilon)$, one should expand $\sinh ^{-l} z$
in powers of $z$ rather than in powers of $\exp(-2z)$.
Fortunately, as $\epsilon\to 0$, the result obtained with a
careful treatment of $\sinh ^{-l} z$ does reduce to the one with
$\sinh ^{-l} z$ expanded naively.

The computations of $A$, $B$, $C$, and $D$ are rather similar in
nature. For illustrations, we will go through the details of
computing $A(\alpha,\beta)$ and its relevant quantities. First of
all, the integration over $v$ can be carried out to give
\begin{eqnarray}\label{Nzv1}
A( \alpha, \beta)
 &=& {\sum}\sqrt{\pi}  K^A_{lmpp'}(q^{\prime 2})
    (\frac{1}{B^{\prime}})^n \frac{({\omega^{\prime}}^2)^{n-l}}
{\Gamma(n-l+\frac{3}{2})}
  \int _0 ^\infty dz \; z^{n-l} \;
    \exp [ {-z\beta} ]
    \;{}_0F_1(n-l+\frac{3}{2};\frac{\alpha^2}{4}z^2),
\end{eqnarray}
where ${}_0F_1$ is the generalized hypergeometric function. We
then perform the $z$ integration which gives
\begin{eqnarray}\label{Nzv3}
A( \alpha, \beta)  &=& {\sum} K^A_{lmpp'}(q^{\prime 2})
   \sqrt{\pi}  (\frac{1}{B^{\prime}})^n
\frac{({\omega^{\prime}}^2)^{n-l}}{(\beta)^{n-l+1}}
\frac{\Gamma(n-l+1)}
{\Gamma(n-l+\frac{3}{2})}\;{}_2F_1(\frac{n-l+1}{2},\frac{n-l+2}{2};
n-l+\frac{3}{2};\frac{\alpha^2}{\beta^2}).
\end{eqnarray}
For later conveniences, it is desirable to disentangle the indices
$n$ and $l$. We do this by replacing $n-l+1$ with $n$ and
replacing the summation ${\sum}$ with the summation ${\sum}^{\prime}\equiv
\sum_{n=1}^{\infty}
\sum_{l=0}^\infty \sum_{m=0}^{\infty}\sum_{p',p=0}^l$. This then
gives rise to
\begin{eqnarray} \label{pipa}
  \Pi_{||,A( \alpha, \beta\, )\,} &=&
- \frac{ \alpha_e m_e^2}{\sqrt{\pi}} {\sum}^{\prime}
K^A_{lmpp'}(q^{\prime 2})
 (B^{\prime})^{1-l}
   (\frac{{\omega^{\prime}}^2}{\beta B^{\prime}})^{n}
\frac{\Gamma(n)}{\Gamma(n+\frac{1}{2})}\;{}_2F_1(\frac{n}{2},\frac{n+1}{2};
n+\frac{1}{2};\frac{\alpha^2}{\beta^2}).
\end{eqnarray}
It is  straightforward to compute the absorption coefficient by
the inverse Mellin transform. We essentially invert the sum rule
given by Eq.~(\ref{mellin}). However, we should remark that
Eq.~(\ref{mellin}) is derived under the assumption $q^2=0$ with
the angle between ${\bf q}$ and ${\bf B}$ kept general. Using this
assumption, one has $q_{\parallel}^2=\omega^2\sin^2\theta$ with
$\theta$ being the angle between ${\bf q}$ and ${\bf B}$. It is
then clear that there is no essential difference for analyzing the
branch cut of $\Pi_{\parallel,\bot}$ either in the complex
$q_{\parallel}^2$-plane or in the complex $\omega^2$-plane. This
is the reason we have chosen $\omega^2$ as the variable to set up
the sum rule in Eq.~(\ref{mellin}). However, in the current case,
$q^2$ is completely general. Hence it is more appropriate to study
the analytic structures of $\Pi_{\parallel,\bot}$ in the complex
$q_{\parallel}^2$-plane. We generalize Eq.~(\ref{mellin}) into
\begin{equation}
{1\over n!}\left({d^n\over d(q_{\parallel}^2)^n}
\hat{\Pi}_{\parallel,\bot}\right)\Big{\vert}_{q_{\parallel}^2=0}
={m_{\parallel,\bot}^{-2n}\over
\pi}\int_{0}^{\infty}du_{\parallel,\bot}\cdot
u_{\parallel,\bot}^{n-1} \cdot \left(\omega
\hat{\kappa}_{\parallel ,\bot}\right),\label{mellin2}
\end{equation}
where $m_{\parallel}^2=4m_e^2$, $m_{\bot}^2=m_e^2(1+\sqrt
{1+2B'})^2$, and
$u_{\parallel,\bot}=m_{\parallel,\bot}^2/q_{\parallel}^2$. Here we
use the notations $\hat{\Pi}_{\parallel,\bot}$ and
$\hat{\kappa}_{\parallel ,\bot}$ to emphasize that the above sum
rule holds for a general angle between ${\bf q}$ and ${\bf B}$. We
also note that the differentiations with respect to
$q_{\parallel}^2$ are understood as acting on
$\bar{\Pi}_{\parallel}(q_{\parallel}^2,q_x^2)$ and
$\bar{\Pi}_{\bot}(q_{\parallel}^2,q_x^2)$ containing inside
$\hat{\Pi}_{\parallel}$ and $\hat{\Pi}_{\bot}$ respectively.
Taking the limit ${\bf q\cdot B}=0$ in the above sum rule, we can
calculate the absorption coefficients $\kappa_{\parallel,\bot}$
through the following inverse Mellin transform:
\begin{equation}\label{trans}
\kappa_{\parallel, \bot}(\omega^2,q_x^2,B) = {1 \over 2i \omega}
\int_C ds F_{\parallel, \bot} (s,q_x^2,B)  (\bar{\omega})^{2s},
\end{equation}
where $F_{\parallel, \bot}(n,q_x^2,B) \equiv
\frac{1}{n!}\frac{d^n}{d({\bar{\omega}}^2)^n}
\Pi_{\parallel,\bot}|_{{\bar{\omega}}^2=0}$ with
$\bar{\omega}=\omega^{\prime}$ for $\parallel$ polarization and
$\bar{\omega}=2\omega^{\prime}/(1+\sqrt{1+2B'})$ for $\bot$
polarization. The integral transform can be easily performed with
the formula
\begin{equation}\label{hyper}
\frac{1}{2\pi i}\int_C ds
x^{-s}\frac{\Gamma(s)}{\Gamma(s+\frac{1}{2})}
\;{}_2F_1(\frac{s}{2},\frac{s+1}{2};
s+\frac{1}{2};z^2)=\frac{1}{\sqrt{\pi}}
\frac{\Theta(1-x+\frac{x^2z^2}{4})}
{\sqrt{1-x+\frac{x^2z^2}{4}}},
\end{equation}
One can derive this formula using the integral representation of
hypergeometric function
${}_2F_1(a,b;c;z)=[\Gamma(c)/(\Gamma(b)\Gamma(c-b))]\int_0^1 dt\;
t^{b-1}(1-t)^{c-b-1}(1-tz)^{-a}$ and the fact that
$\frac{1}{2\pi i}\int_C ds x^{-s}u^{s-1}=\delta(x-u)$.
Therefore, we arrive at
\begin{eqnarray}\label{abso}
    \kappa_{||,A( \alpha, \beta\, )\,}
&= &-  \frac{\alpha_e m_e B'}{2\omega^{\prime}}
        {\sum}^{\prime \prime}
{K^A_{lmpp'}(q^{\prime 2}) \Theta\left((1-\frac{\beta
B^{\prime}}{{\omega^{\prime}}^2}) +(\frac{\alpha
B^{\prime}}{2\omega^{\prime 2}})^2\right)\over B^{\prime l}
\sqrt{(1-\frac{\beta B^{\prime}}{{\omega^{\prime}}^2})
          +(\frac{\alpha B^{\prime}}{{2 \omega^{\prime}}^2})^2}},
\end{eqnarray}
where ${\sum}^{\prime \prime}\equiv \sum_{l=0}^\infty
\sum_{m=0}^{\infty}\sum_{p',p=0}^l$. To understand the structure
of $\kappa_{\parallel,A}$, we rewrite the denominator of the above
equation as
\begin{equation}\label{sing}
\sqrt{(1-\frac{\beta B^{\prime}}{\omega^{\prime 2}})+
(\frac{\alpha B^{\prime}}{2\omega^{\prime
2}})^2}=\frac{1}{\omega^{\prime}} \sqrt{\omega^{\prime
2}-1-(l_1+l_2)B^{\prime}+(l_1-l_2)^2\frac{B^{\prime 2}}
{4\omega^{\prime 2}}},
\end{equation}
with $(\beta + \alpha)B'=1+2l_1 B'$, and $(\beta -\alpha)B'=1+2l_2
B'$. By a simple kinematic analysis, one can show that the R.H.S.
of Eq. (\ref{sing}) is just ${\vert p_{1,z}\vert}/m_e
{\omega^{\prime}}\equiv {\vert p_{2,z}\vert} /m_e
{\omega^{\prime}}$ where $p_{1,z}$ and $p_{2,z}$ are the
$z$-direction momenta of $e^-$ and $e^+$ respectively; while $l_1$
and $l_2$ are the Landau levels occupied by $e^{-}$ and $e^+$. The
pair-production threshold corresponds to $p_{1,z}\equiv
p_{2,z}=0$. This threshold behavior is seen explicitly from the
step function in the expression for $\kappa_{\parallel, A}$. It is
worthwhile to point out that the variables $\alpha$'s and
$\beta$'s, appearing first in Eqs.~(\ref{component}) and
(\ref{pi_perp}), do have physical meanings. Taking
$A(\alpha,\beta)$ as an example, different values of $\alpha$ and
$\beta$ imply a different radius of convergence for the infinite
series in $\omega^{\prime}$ (see Eq.~(\ref{Nzv1})). The absorptive
part of $A(\alpha,\beta)$ emerges once the scaled photon energy,
$\omega^{\prime}$, becomes greater than the radius of convergence.
This is reflected by the inverse Mellin transform given by Eq.
(\ref{hyper}).

It is desirable to further simplify Eq.~(\ref{abso}). We first
perform the summation over the index $l$ in $K^A_{lmpp'}$:
\begin{eqnarray}
\sum_{l=0}^{\infty}\sum_{p,p'=0}^l\frac{
x^l\Gamma(l+m+1)}{(l-p)!(l-p')!} &=& \sum_{p,p'=0}^{\infty}
x^{\bar{p}}e^x\Gamma(\hat{p}+m+1)L_{\hat{p}+m}
^{\bar{p}-\hat{p}}(-x),
\end{eqnarray}
where $x=-2q^{\prime 2}/B'\equiv -q_x^2/(2eB)$, $\bar{p}={\rm max}
\{p,p'\}$, $\hat{p}={\rm min}\{p,p'\}$, and $L_{\hat{p}+m}
^{\bar{p}-\hat{p}}$ is the Laguerre polynomial. To derive this
equation, we have used the relations ${}_1F_1(\alpha;\beta;x)=e^x
{}_1F_1(\beta-\alpha;\beta;-x)$ and
$L_n^{\alpha}(x)={}_1F_1(-n;\alpha+1;x)\times C_n^{n+\alpha}$. At
this juncture, the summation ${\sum}^{\prime \prime}$ in
Eq.~(\ref{abso}) has already turned into
$\sum_{p,p'=0}^{\infty}\sum_{m=0}^{\infty}$. To proceed, we recall
that the actual values for $\alpha$ and $\beta$ in this case are
$\alpha_a=p'-p+1$ and $\beta_a=p+p'+2m+1+1/B'$, which correspond
to the Landau levels $l_1=p'+m+1$ and $l_2=p+m$. Therefore the
above summation is equivalent to
$\sum_{l_1=1}^{\infty}\sum_{l_2=0}^{\infty}\sum_{m=0}^{\hat{l}}$
with $\hat{l}={\rm min}\{l_1-1,l_2\}$. By collecting the relevant
terms in $\kappa_{\parallel,A(\alpha_a,\beta_a)}$, we can further
sum over the index $m$:
\begin{equation}
\sum_{m=0}^{\hat{l}}x^{-m}\frac{1}{\Gamma(\bar{l}-m+1)}
\frac{1}{\Gamma(\hat{l}-m+1)}\frac{1}{\Gamma(m+1)}=
\frac{x^{-\hat{l}}}{\Gamma(\bar{l}+1)}L_{\hat{l}}^{\bar{l}-\hat
{l}}(-x),
\end{equation}
where $\bar{l}={\rm max}\{l_1-1,l_2\}$. With the above summations
over $l$ and $m$, we arrive at
\begin{eqnarray}
    \kappa_{\parallel,A}\label{kaA}
&= &\frac{\alpha_e m_e B'}{\omega^{\prime}}
        \sum_{l_1=1}^{\infty}\sum_{l_2=0}^{\infty}
{{\bf T}^A_{l_1 l_2}(x)
\Theta\left((1-\frac{\beta B^{\prime}}{{\omega^{\prime}}^2})
+(\frac{\alpha B^{\prime}}{2\omega^{\prime 2}})^2\right)\over
\sqrt{(1-\frac{\beta B^{\prime}}{{\omega^{\prime}}^2})
          +(\frac{\alpha B^{\prime}}{{2 \omega^{\prime}}^2})^2}},
\end{eqnarray}
with\footnote{The absorption coefficient is not written in a
symmetrized form for saving the space. Nevertheless, the
symmetrization with respect to $l_1$ and $l_2$ can be easily done
as one wishes.}
\begin{equation}
{\bf T}^A_{l_1
l_2}(x)=(-1)^{1+r_A}x^{r_A}e^x\frac{\Gamma(\lambda_A+1)}
{\Gamma(\lambda_A+r_A+1)}L_{\lambda_A}^{r_A}(-x)L_{\lambda_A}^{r_A}(-x),
\end{equation}
where $r_A\equiv \bar{l}-\hat{l}=|l_1-l_2-1|$, and
$\lambda_A\equiv \hat{l}=(l_1+l_2-|l_1-l_2-1|-1)/2$. The type $B$
contributions to $\kappa_{\parallel}$ arise from
$B(\alpha_b,\beta_b)$ and $B(\alpha_b^{\prime},\beta_b^{\prime})$,
We write down the results without repeating the details:
\begin{eqnarray}\label{kaB}
  \kappa_{\parallel,B}&=&\frac{\alpha_e m_e B^{\prime 2}}{2\omega^{\prime 3}}
        \sum_{l_1=0}^{\infty}\sum_{l_2=0}^{\infty}
{{\bf T}^{B}_{ l_1 l_2}(x) \left(\beta-\frac{\alpha^2
B'}{2\omega^{\prime 2}}\right) \Theta\left((1-\frac{\beta
B^{\prime}}{{\omega^{\prime}}^2}) +(\frac{\alpha
B^{\prime}}{2\omega^{\prime 2}})^2\right)\over
\sqrt{(1-\frac{\beta B^{\prime}}{{\omega^{\prime}}^2})
          +(\frac{\alpha B^{\prime}}{{2 \omega^{\prime}}^2})^2}},
\end{eqnarray}
where
\begin{equation}
{\bf T}^{B}_{l_1
l_2}(x)=(-1)^{r_B}x^{r_B}e^x\frac{\Gamma(\lambda_B+1)}
{\Gamma(\lambda_B+r_B+1)}L_{\lambda_B}^{r_B}(-x)L_{\lambda_B}^{r_B}(-x)+
(\lambda_B\to \lambda_B-1),
\end{equation}
with $r_B=|l_1-l_2|$, $\lambda_B=(l_1+l_2-|l_1-l_2|)/2$. We note
that the contribution by $B(\alpha_b^{\prime},\beta_b^{\prime})$
is incorporated by the replacement $\lambda_B\to \lambda_B-1$ in
the above equation. Finally, the type $C$ contributions are given
by
\begin{eqnarray}
    \kappa_{\parallel,C}\label{kaC}
&= &\frac{\alpha_e m_e B^{\prime 2}}{2\omega^{\prime 3}}
        \sum_{l_1=1}^{\infty}\sum_{l_2=0}^{\infty}
{{\bf T}^C_{l_1 l_2}(x)
\Theta\left((1-\frac{\beta B^{\prime}}{{\omega^{\prime}}^2})
+(\frac{\alpha B^{\prime}}{2\omega^{\prime 2}})^2\right)\over
\sqrt{(1-\frac{\beta B^{\prime}}{{\omega^{\prime}}^2})
          +(\frac{\alpha B^{\prime}}{{2 \omega^{\prime}}^2})^2}},
\end{eqnarray}
where
\begin{equation}
{\bf T}^C_{l_1
l_2}(x)=(-1)^{r_C}x^{r_C}e^x(l_1-l_2)\frac{\Gamma(\lambda_C+2)}
{\Gamma(\lambda_C+r_C+1)}\left(L_{\lambda_C}^{r_C}(-x)L_{\lambda_C}^{r_C}(-x)
-L_{\lambda_C+1}^{r_C-1}(-x)L_{\lambda_C-1}^{r_C+1}(-x)\right)+(\lambda_C\to
\lambda_C-1),
\end{equation}
with $r_C=r_A$ and $\lambda_C=\lambda_A$. We like to emphasize
that, in $\kappa_{\parallel}$, the relation between
$\omega^{\prime 2}$ and $q^{\prime 2}$ is still kept general. It
is the dispersive part of $\Pi_{\parallel}$ that determines the
relation between $\omega^{\prime 2}$ and $q^{\prime 2}$ . The
dispersive part can be calculated using the {\it Kramers-Kronig}
relation:
\begin{equation}
{\rm Re}\bar{\Pi}_{\parallel}(\omega^{\prime 2},q^{\prime 2})=
\frac{P}{\pi}\int_{1}^{\infty}dt\frac{{\rm Im}
\bar{\Pi}_{\parallel}(t,q^{\prime 2})}{t-\omega^{\prime 2}},
\end{equation}
where $P$ stands for evaluating the principle part of the
integral, while the dimensionless function
$\bar{\Pi}_{\parallel}(\omega^{\prime 2},q^{\prime 2})$ is related
to $\Pi_{\parallel}(\omega^2,q_x^2)$ by
$\Pi_{\parallel}(\omega^2,q_x^2)=-(\alpha_e\omega^2/ 4\pi)
\bar{\Pi}_{\parallel}(\omega^{\prime 2},q^{\prime 2})$, as has
been shown right above Eq.~(\ref{pibar}). It is well known that
${\rm Re}\Pi_{\parallel}$ is relevant to the photon refractive
index $n_{\parallel}(\omega^{\prime})$ through the equation
$q^2+{\rm Re}\Pi_{\parallel}=0$. Hence
\begin{equation}\label{refrac}
 n_{\parallel}(\omega^{\prime})=1+\frac{P}{4m_e\pi}\int_1^{\infty}
 dt \frac{\kappa_{\parallel}(t)}{\sqrt{t}(t-\omega^{\prime 2})}.
\end{equation}
For illustrations, we evaluate $n_{\parallel}$ for a below
threshold energy $\omega^{\prime}<1$ and a super-critical magnetic
field $B\gg B_c$. To the leading order in $B_c/B$, it suffices to
include the lowest Landau-level contribution on the R.H.S. of the
above equation. Such a contribution is contained in
$\kappa_{\parallel,B}$, i.e.,
\begin{equation}\label{ground}
\kappa_{\parallel,B}^{l_1=l_2=0}=\frac{\alpha_e m_e
B'}{2\omega^{\prime 2} \sqrt{\omega^{\prime
2}-1}}\exp(-{2q^{\prime 2}\over B'}),
\end{equation}
Combining Eqs. (\ref{refrac}) and (\ref{ground}), we find
\begin{equation}
n_{\parallel}(\omega^{\prime})=1+\frac{\alpha_e
B'}{4\pi\omega^{\prime 2}} \left(\frac{1}{\sqrt{\omega^{\prime
2}(1-\omega^{\prime 2})}} \arctan\sqrt{\frac{\omega^{\prime
2}}{1-\omega^{\prime 2}}} -1\right),
\end{equation}
which agrees with the previous result\cite{shabad}.

We now turn our attentions to the absorption
coefficient $\kappa_{\bot}$.
It is useful to rewrite Eq. (\ref{trans}) as
\begin{equation}
\kappa_{\bot}(\omega,q_x^2,B) = {1 \over 2i \omega} \int_C ds
{\tilde F}_{\bot} (s,q_x^2,B)  (\omega^{\prime})^{2s},
\end{equation}
where ${\tilde F}_{\bot}(n,q_x^2,B) \equiv
\frac{1}{n!}\frac{d^n}{d({\omega^{\prime}}^2)^n}
\Pi_{\bot}|_{\omega^{\prime 2}=0}$. With $\kappa_{\bot}$ written
in this form, we can easily compute $\kappa_{\bot}$ using the
results from $\kappa_{\parallel}$. Comparing Eqs.
(\ref{component}) and (\ref{pi_perp}), we can easily show that
$\kappa_{\bot,A}=- (q^{\prime 2}/ \omega^{\prime 2})
\kappa_{\parallel,A}$, and $\kappa_{\bot,C}=- (q^{\prime 2}/
\omega^{\prime 2}) \kappa_{\parallel,C}$. Finally, it is slightly
involved to compute the type $D$ contribution. We obtain
\begin{eqnarray}
    \kappa_{\bot,D}\label{keD}
&= &\frac{\alpha_e m_e B'}{2\omega^{\prime}}(\frac{q^{\prime 2}}
{\omega^{\prime 2}})
        \sum_{l_1=1}^{\infty}\sum_{l_2=1}^{\infty}
{{\bf T}^D_{l_1 l_2}(x)
\Theta\left((1-\frac{\beta B^{\prime}}{{\omega^{\prime}}^2})
+(\frac{\alpha B^{\prime}}{2\omega^{\prime 2}})^2\right)\over
\sqrt{(1-\frac{\beta B^{\prime}}{{\omega^{\prime}}^2})
          +(\frac{\alpha B^{\prime}}{{2 \omega^{\prime}}^2})^2}},
\end{eqnarray}
with
\begin{equation}
{\bf T}^D_{l_1
l_2}(x)=8(-1)^{1+r_D}x^{r_D-1}e^x\frac{\Gamma(\lambda_D+1)}
{\Gamma(\lambda_D+r_D)}L_{\lambda_D-1}^{r_D+1}(-x)L_{\lambda_D}^{r_D-1}(-x),
\end{equation}
with $r_D=r_B$ and $\lambda_D=\lambda_B$.
Similar to the case of $\parallel$ polarization,
the dispersive part of $\Pi_{\bot}$ can also be calculated using
the {\it Kramers-Kronig} relation. Since the technique is identical, we will not
dwell upon this issue again.

We have compared our absorption coefficients with those obtained
by squaring the $\gamma\to e^+e^-$ amplitude directly\footnote{We
make comparisons with the most updated results in Ref.
\cite{DH}.}. Our results reduce to that of Daugherty and Harding
in the special limit $q^2=0$ which they have assumed. This is
similar to what Shabad has demonstrated in Ref. \cite{prep} as he
compared his result with that of Klepikov \cite{kle} in the
above-mentioned limit for $q^2$. Our approach differs from that of
Refs. \cite{shabad,prep} in that Shabad performs the calculation
in the beyond-threshold energy where the algebraic manipulations
are rather involved and precautions are required, whereas we take
the advantage of inverse Mellin transform which permits us to
calculate the polarization function near the zero energy with a
convenient energy expansion. We wish to stress again that our
approach is physically intuitive. We have written
$\Pi_{\parallel,\bot}$ as a multiple series in $\omega$ (photon
energy), $\alpha$ and $\beta$. It is possible to show
that\cite{kl}, for fixed values of $\alpha$ and $\beta$, the
series in $\omega$ begins to diverge at the threshold energy
$\omega_{\rm th}\equiv m_e(\sqrt{1+2l_1 B'}+\sqrt{1+2l_2 B'})$
where the Landau levels $l_1$ and $l_2$ are given by
$l_1=(\beta+\alpha-1/B')/2$ and $l_2=(\beta-\alpha-1/B')/2$. The
divergence of the energy series beginning at $\omega=\omega_{\rm
th}$ implies the existence of absorptive part beyond this point.
We are able to compute the absorptive part as well as the
dispersive part at any energy with the help of inverse Mellin
transform and {\it Kramers-Kronig} relation.

In closing, we have developed an integral-transform technique to
compute the absorptive part of photon polarization function in a
background magnetic field, while the dispersive part can be
obtained via the {\it Kramers-Kronig} relation. Although we have
chosen a special case ${\bf q}\cdot {\bf B}=0$ for convenience,
the result for a general angle between ${\bf q}$ and ${\bf B}$ is
easy to infer. Using the relations
$\hat{\Pi}_{\parallel}(\omega^2,q_x^2,\theta)=
-(\alpha_e\omega^2\sin^2\theta/ 4\pi)
\bar{\Pi}_{\parallel}(q_{\parallel}^2,q_x^2)$,
$\hat{\Pi}_{\bot}(\omega^2,q_x^2,\theta)=-(\alpha_e q_x^2/ 4\pi)
\bar{\Pi}_{\bot}(q_{\parallel}^2,q_x^2)$, and
$\hat{\kappa}_{\parallel,\bot} = {\rm
Im}\hat{\Pi}_{\parallel,\bot}/ \omega$, it is clear that
$\hat{\kappa}_{\parallel}(\omega^2, q_x^2, \theta)= -(\alpha_e
\omega \sin^2 \theta/ 4\pi){\rm
Im}\bar{\Pi}_{\parallel}(q_{\parallel}^2, q_x^2)$, and
$\hat{\kappa}_{\bot}(\omega^2, q_x^2, \theta)= -(\alpha_e q_x^2 /
4\pi\omega){\rm Im}\bar{\Pi}_{\bot}(q_{\parallel}^2, q_x^2)$. For
$\theta=\pi/2$, i.e., ${\bf q}\cdot {\bf B}=0$,
$\bar{\kappa}_{\parallel}\to \kappa_{\parallel}=-(\alpha_e \omega
/ 4\pi){\rm Im}\bar{\Pi}_{\parallel}(\omega^2, q_x^2)$ and
$\bar{\kappa}_{\bot}\to \kappa_{\bot}=-(\alpha_e q_x^2 /
4\pi\omega){\rm Im}\bar{\Pi}_{\bot}(\omega^2, q_x^2)$. Hence, for
example, by comparing the above expressions for
$\hat{\kappa}_{\parallel}$ and $\kappa_{\parallel}$, one realizes
that $\hat{\kappa}_{\parallel}$ can be inferred from
$\kappa_{\parallel}$ by first dividing the latter by $\omega$,
then replacing the variable $\omega^2$ by $q_{\parallel}^2$, and
finally multiplying the entire expression by $\omega\sin^2\theta$.
The procedure of obtaining $\hat{\kappa}_{\bot}$ from
$\kappa_{\bot}$ is also straightforward. There are other
generalizations to the current work. For example, one may analyze
the photon polarization function in a general background
electromagnetic field, or study other current-current correlation
functions under the same external condition. To our knowledge, the
vector-axial vector correlation function relevant to weak
interaction processes has not been analyzed as detailed as the
fashion presented in this work. We shall report the results of
such analysis as well as some technical details omitted in this
letter in a forthcoming publication\cite{kl}.

\acknowledgments We thank S.-C. Lee for his comments which lead to
this investigation. This work is supported in part by the National
Science Council under grant numbers NSC89-2112-M009-043 and
NSC89-2112-M009-041.

\end{document}